\documentclass[aps,prb,showpacs,superscriptaddress,amsmath,amssymb,twocolumn]{revtex4}

\usepackage{mathptmx} 
\usepackage{helvet} 
\usepackage{courier} 
\usepackage{type1cm} 
\usepackage{makeidx} 
\usepackage{graphicx} 
\usepackage{epstopdf} 
\usepackage{verbatim} 
\usepackage{amsmath}
\usepackage{calc}
\usepackage{enumitem}
\usepackage{amssymb}
\usepackage{latexsym}
\usepackage{bm}
\usepackage{color}

\makeindex 

\usepackage[
	colorlinks,
	linkcolor=blue,
	anchorcolor=blue,
	citecolor=blue,
	urlcolor=blue,
	menucolor=blue,
	filecolor=blue
]{hyperref}

\begin{document}

\title{Charged domain walls in BaTiO$_3$ crystals emerging from superdomain boundaries}



\author{Petr S. Bednyakov}
\author{Ji\v{r}\'{\i} Hlinka}
\affiliation{FZU-Institute of Physics$,$ The Czech Academy of Sciences$,$ Na Slovance 2$,$ 18221 Praha 8$,$ Czech Republic}
\email{bednyakov@fzu.cz, hlinka@fzu.cz}



\date{\today}

\begin{abstract}
Previous experiments with BaTiO$_3$ single crystals have shown that application of the electric field in the vicinity of the ferroelectric phase transition can be used to introduce peculiar persisting ferroelectric domain walls, accompanied by the compensating charge in the form of two-dimensional electron gas.
The present {\it in-situ} optical observations of such electric poling process reveal  formation of a transient coexistence of the cubic and ferroelectric phases, the latter one being broken into multiple martensitic superdomains, separated by superdomain walls.
%
It is revealed that as the transient superdomains convert into the regular ferroelectric domains, the superdomain boundaries transform into the desired charged domain walls.
In order to assign the observed transient domain patterns,  to understand the shapes of the observed ferrolectric precipitates and their agglomerates as well as to provide the overall interpretation of the observed domain formation process, the implications of the mechanical compatibility of the coexisting superdomain states is derived in the framework of the Wechsler-Lieberman-Read theory.
The results also suggest that the transport of the compensating charge carriers towards the final charged domain wall location is directly associated with the electric conductivity and interlinked motion and growth of the superdomain walls and phase fronts. 
%
 \end{abstract}

\maketitle

\section{Introduction}

%
Ferroelectric charged domain walls  have recently attracted lots interest due to their ultimate nanoscale thickness combined with promising charge transport properties \cite{sluka2013free, werner2017large,  bednyakov2018physics}.
Conditions of the stability of the charged domain
walls in strong ferroelectrics like BaTiO$_3$ has been discussed and clarified in terms of phenomenological and microscopic theory\cite{sturman2015quantum,sturman2017charged,sturman2018ion,gureev2011head,vul1973encountering}.
Experimentally, it has been shown that charged domain walls can be engineered with a controlled density\cite{bednyakov2015formation, bednyakov2016free}, written locally with scanning probe tools\cite{crassous2015polarization} or using patterned electrodes \cite{jiang2017temporary}, and also  repeatedly reconfigured \cite{maksymovych2012tunable, schroder2012conducting, li2016giant}.
The possibility to address the conductivity of the domain wall has been even exploited in various device proposals   \cite{sharma2017nonvolatile, li2016giant, mundy2017functional, sluka2015patent, jiang2017temporary}. 

At ambient temperature, ferroelectric BaTiO$_3$ crystal has tetragonal crystal symmetry.  
Its 6 primary ferroelectric domain states are related to the $m\bar3m > 4mm$ macroscopic symmetry breaking (species No.\,198 of Ref.\,\onlinecite{hlinka2016symmetry}).
When disregarding the small angles associated with the elastic matching of domain states and adopting to the pseudocubic  crystallographic notation, we can state that the polarization falls along one of the six $\{001\}$ directions and  that  180 degree and 90-degree ferroelectric domain wall can be formed.
The 90-degree walls are particularly attractive for domain engineering because they are also ferroelastic and can be induced by the frustrative poling that favors more than one ferroelectric domain state. 
The existence of charged 90-degree domain walls always require almost perfect screening of the huge bound polarization charge at such a wall.
The usual BaTiO$_3$ crystals are slightly n-doped semiconductors, so that the head-to-head charged domain walls can be compensated by the naturally present excess electrons\cite{sturman2018ion}.
The concentration of the compensation charge is so strikingly high that even the spontaneous formation of the 
two-dimensional electron gas at this 90-degree head-to-head domain wall have been observed in  BaTiO$_3$ crystal\cite{sluka2013free,beccard2022nanoscale}.


%
%

%

\begin{figure}[h]
\centering
\includegraphics[width=0.95\columnwidth]{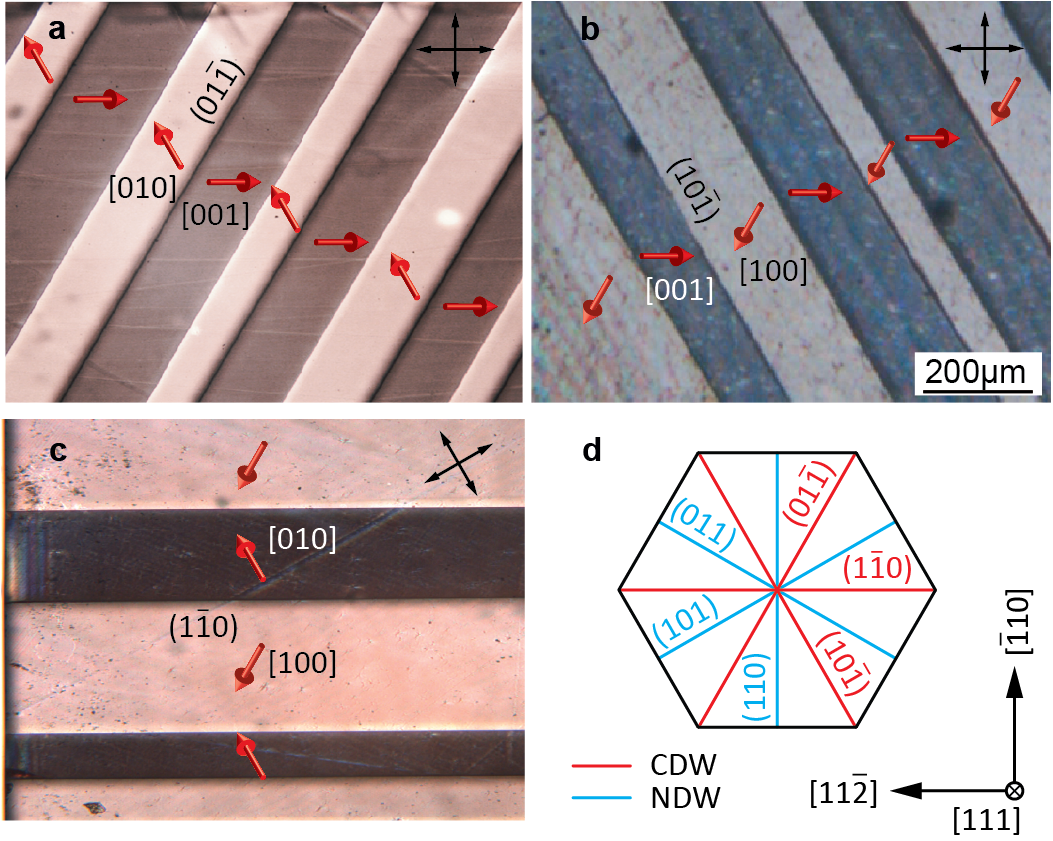}
\caption{ 
  Charged domain walls in (111)-oriented BaTiO$_3$ crystal plates.
   Panels (a), (b) and (c) shows $(01\bar{1})$, $(10\bar{1})$ and $(1\bar{1}0)$ systems of charged domain walls.  
  These ex-situ optical micrographs are taken in transmission mode,
  the viewing direction is that of the formerly applied [$111$]-directed poling field.  
 Red arrows stand for the experimentally inferred polarization in the primary domain states favored by the poling electric field, while the
   the polarizer-analyzer orientation is indicated by the crossed directors.
Panel (d) shows how the mechanically compatible neutral (NDW) and charged (CDW) primary ferroelastic domain walls can be terminated on the (111) surface. 
}
\label{fig1}
\end{figure}

 %
 %
 %
 
 %
It is worth noting that head-to-head and tail-to-tail ferroelastic domains walls form parallel planes alternating in the crystal (see Fig.\,\ref{fig1}). 
Consequently, the
positively charged carriers, e.g. oxygen vacancies or other ionized mobile defects or electron holes, also need to be present in the crystal in order to form a stable domain pattern with multiple domain walls. 
 The sufficient amount of mobile positive charge carriers  have been secured using two alternative strategies.
 In the first approach,  crystals containing a sufficient amount of oxygen vacancies were used\cite{bednyakov2015formation}.
 In the second approach,  the poling was performed under the UV light illumination\cite{sluka2013free}, suggesting that both photoexcited electrons and holes participated in screening of the polarization bound charge\cite{bednyakov2016free}.

 Strictly speaking, the process of charged domain wall creation in BaTiO$_3$ requires not only the presence of charge carriers but also their efficient spatial redistribution.
 For example, it has been recently emphasized that formation of the charged domain walls in the polarization switching process is much more understandable when such are conductive and connected to the outer electrode\cite{sturman2022ferroelectric}.
However, the exact mechanism of the 90-degree charge domain pattern formation process was not fully understood yet. 
 In particular, there is no obvious reason for an organized charge carrier separation before some sort of bound charge modulation is formed in the crystal.
 Likewise, the idea that the macroscopic domain patterns could be initially nucleated without the full compensation and only then stabilized by the carriers of both signs appears unrealistic when considering plausible charge carrier drift velocities and the micron-sized distances between the charged domain walls.

 In this work, we draw the attention to the fact that the 90-degree charged domain patterns were so far created in bulk crystals of BaTiO$_3$ by the process of frustrative electric poling in the vicinity of  phase transitions\cite{bednyakov2015formation,bednyakov2016free}. With this in mind,
 we have been revisiting the problem of the  charged domain walls creation by careful  {\it in-situ} optical investigations of the early stage of their formation near
 the cubic to tetragonal phase transition\cite{bednyakov2015formation}.
 We have realized that the charged domain walls are actually originating from the charged superdomain walls, separating nanotwinned martensitic plates, here also called superdomains, that are formed in the process of the nucleation of the ferroelectric phase itself.
 The optical observations are complemented by an in-detph theoretical analysis of these superdomain structures and walls. 

The paper is organized as follows. 
In Section II., we present  the main experimental observation that are documented in Figs.\,\ref{fig1}, \ref{fig2} and \ref{fig3}. 
Then, in Section III. we expose the result of the theoretical analysis.
The part III.A summarizes the known results of the Wechsler-Lieberman-Read (WLR)  martensite theory of phase fronts in BaTiO$_3$.
Among others, we clarify there what exactly we mean by superdomains.
The rest of the Section III. describes  various properties of the superdomain walls and superdomain precipitates and the expected consequences for the adopted geometry of the frustrative poling with [111] oriented field as we have derived them
from the mechanical compatibility conditions.
The main results are given in Tables I. and II. and in the summarizing Figure\,\ref{fig4}. 
In Section IV., the theoretical background is used to analyze and interpret the principal results of the Section II.
Next, we show that the so far derived conclusions are in agreement with some additional observations and several proposals for further investigation are discussed. Finally, the paper is concluded by considerations about the role of possibly conductive phase fronts.

\section{Principal observations}

%
%

\begin{figure*}[ht]
\centering
\includegraphics[width=1.95\columnwidth]{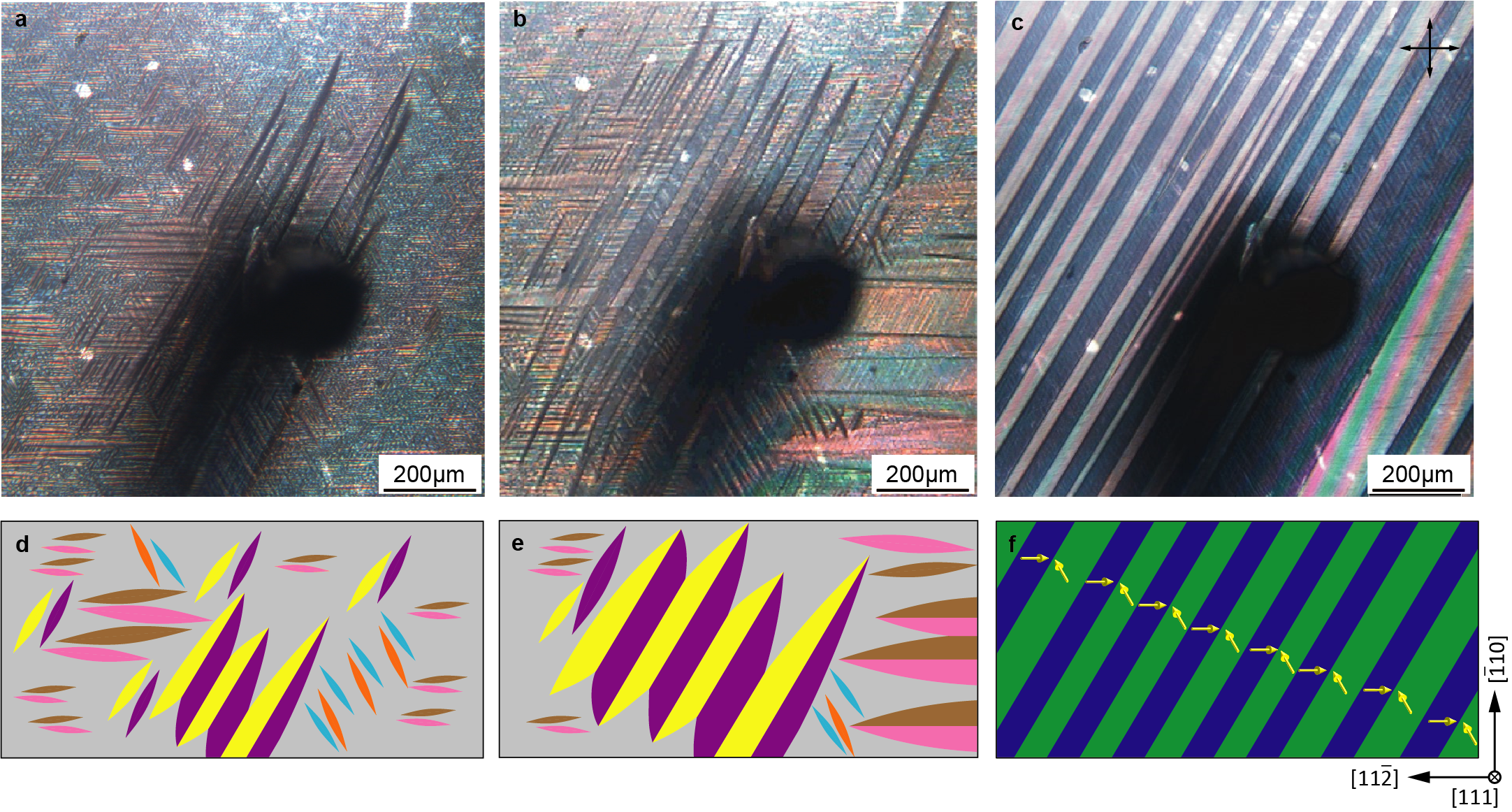}
\caption{
Optical micrographs showing the charged domain wall formation process in
(111)-oriented crystal plate under the electric field applied along the viewing direction.
Panel (a)  suggests nucleation of the ferroelectric precipitates in the form of thin platelets embedded in the optically isotropic paralectric matrix. 
 In the viewing direction the platelet precipitates appear as needle-shaped leafs oriented similarly as the traces of the charged domain walls in Fig.\,\ref{fig1}.
Panel (b) reveals that the  needle-shaped leafs have two blades, each of them decorated with a fine stripe pattern, and that the precipitates become organized in a  pattern with common zigzag phase fronts.
Panel (c) shows that the fine stripe pattern is fading out and 
the that stable charged domain walls are formed at the central lines of the needle-shape leaves. 
The panels (d,e,f) suggest schematic representation of the corresponding observations of panels (a,b,c). 
Black spot in the center of images (a,b,c) is due to an electric contact on the top electrode; polarizer-analyzer orientation is indicated in upper right corner of panel (c).
}
\label{fig2}
\end{figure*}

\begin{figure}[h]
\centering
\includegraphics[width=0.95\columnwidth]{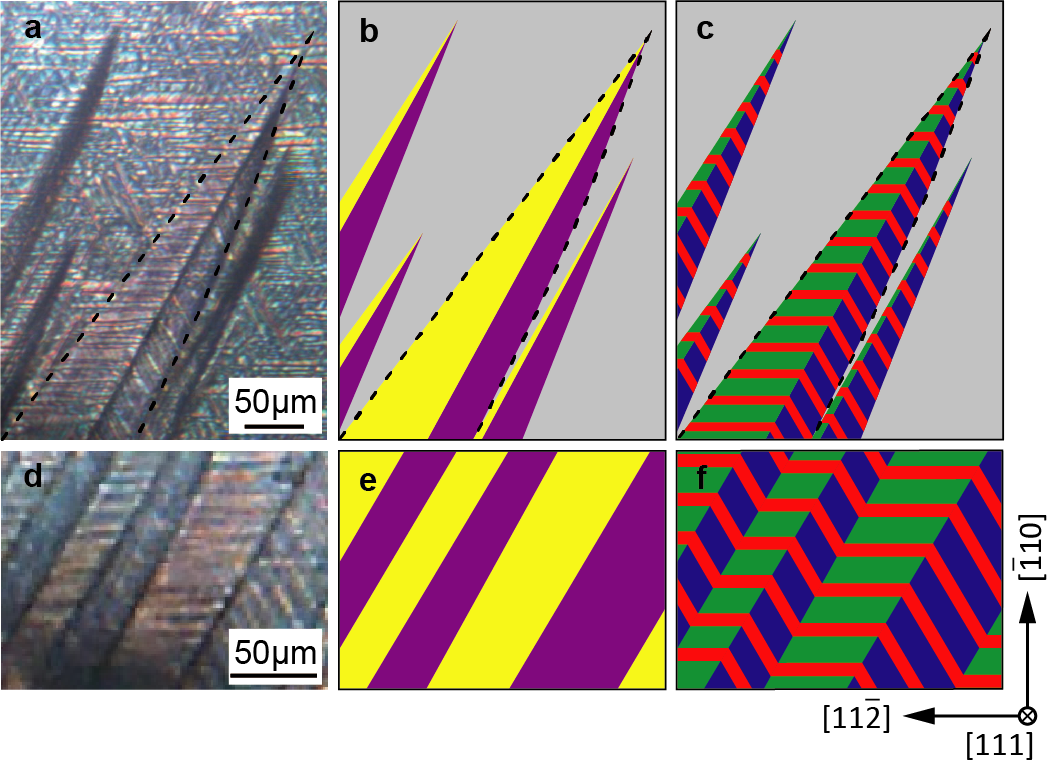}
\caption{
Details of the transient domain structure shown in Fig.\,\ref{fig2}. 
Top panels show awl-shaped ferroelectric precipitates surrounded by the paraelectric phase. 
The region captured by the optical micrograph in panel (a) is graphically interpreted in terms of Zx and Yx superdomains in panel (b).
The fine-scale stripes due to the primary ferroelectric domains within these superdomains  form the resulting pattern of panel (c), reminding of awl-shaped leaves with two blades.
Dashed lines highlight phase boundaries of one such precipitate. 
By comparison with the Fig.\,\ref{fig4} explained in Section III., the junction of the two blades can be identified as a
type I.A superdomain boundary.
The bottom panels (d-f) show the corresponding information for an area fully covered by the ferroelectric phase.
Note that the  assignment of the strain state in superdomains was done using the predictions of the Section III., summarized in Fig.\,\ref{fig4}.
%
%
The polarization content of the individual primary domains can be inferred from a more subtle arguments given in section\,IV (see also Fig.\,\ref{fig3bis} and color code of Fig.\,\ref{fig4}).
%
%
%
%
%
}
\label{fig3}
\end{figure}

The main results of the present experiments are summarized in Figs.\,\ref{fig1}-\ref{fig3}.
The samples used in all investigations were (111)-oriented platelets.
Formation of straight charged domain walls was most conveniently achieved in slightly off-stoichiometric BaTiO$_3$ single crystals with conductivity of the order of 0.1-1 S/m. 
%
%
Similarly as is the Ref.\,\onlinecite{bednyakov2015formation}, these  charged walls were induced by frustrative poling method
across the cubic to tetragonal phase transition.
More precisely, the [111]-oriented dc electric field of about 1\,kV/mm field was applied several K above the phase transition and then the sample was cooled down through the transition while the field was kept on. 
After cooling the sample down to the ambient temperature and switching off the field, the contacts were removed and the resulting domain patterns were thoroughly investigated ex-situ.
We have confirmed that most of the charged domain walls grown in this way were stable at least over several months.

%

%
Typically, such samples have shown an array of  parallel head-to-head and tail-to-tail domain walls separated by distances of about 100 microns (see Fig.\,\ref{fig1}).
Interpretation of the results shown in Fig.\,\ref{fig1} is rather straightforward.
Polarization vectors were identified by the orientation of the optical indicatrix and the light extinction at the domain wall in the non-polarized light following the procedure described in Ref.\,\onlinecite{bednyakov2015formation}.
Distinction between the head-to-head and tail-to-tail could be easily made based on the observation of the dark and light line at the domain wall location\cite{bednyakov2015formation}.
Comparison with Fig.\,\ref{fig1}d confirms that crystallographic orientations of these charged domain walls agree with the three theoretically expected orientations of primary charged domain walls, as discussed for example in Refs.\,\onlinecite{fousek1969orientation, bednyakov2015formation}.

Repeating the same poling procedure using thin transparent electrodes enabled us to observe also the transient domain structures formed at the early stage of the charge domain wall formation. 
This experiment, essential for the present paper, is reported in Fig.\,\ref{fig2}.
It was carried out with a similar sample as that of Fig.\,\ref{fig1}. 
The precipitation started at temperature $T_{\rm start}$ which was about 402\,K.
Interestingly, we have observed that the transformation started by a simultaneous nucleation of many thin, lentil-shaped precipitates, appearing within the bulk of the optically homogeneous and isotropic paraelectric cubic phase.
The boundaries of the precipitates perform back and forth jerky motion, typical for the dynamics of ferroelastic domain walls.
The image of Fig.\,\ref{fig2}a shows these precipitates at a temperature of about about $T_{\rm start}-0.5$\,K. 
The cross section of the precipitate with the surface appears as a needle-shaped leaf.
 The individual needles appear to be organized in clusters, suggesting a systematic preference or a direct interaction among the parallel precipitates.
Fig.\,\ref{fig2}b shows the same area at a later stage, corresponding to the temperature of about $T_{\rm start}-1$\,K.
One can see that precipitates formed  compact fan-like domain patterns delimited by two zig-zag boundaries (see Fig.\,\ref{fig2}b and its schematic interpretation in Fig.\,\ref{fig2}e).
Originally, three competing systems of parallel needle-shaped leaves were coexisting in the sample, each corresponding to
of one of the charged domain wall orientations shown in Fig.\,1. 
Progressively, only one of the three systems of precipitates extended over the whole crystal (Fig.\,\ref{fig2}c).
%

A closer look to one of the  thicker precipitates shown in Fig.\,\ref{fig2}b allows to discern a central line, dividing each  needle-shaped leaf into two adjacent blades.
Let us note that for the given orientation of the crossed polarizers, one type of the blades is systematically darker, while the other is barely seen.
The relevant detail is thus enlarged in Fig.\,\ref{fig3}a and schematized in Fig.\,\ref{fig3}b.
There one can also see that each of the adjacent blade is decorated inside with a differently oriented fine-stripe pattern.
From the analysis of this fine-stripe pattern and the overall geometry of the precipitates detailed in the next section, we infer that the individual blades of the precipitates correspond to distinct martensitic superdomains.
 In Fig.\,\ref{fig3}b,
these two blades are shaded by different colors, using the superdomain color code defined in Fig.\,\ref{fig4}a. 
Similar but more schematic graphical interpretation of Figs.\,\ref{fig2}a and \ref{fig2}b is given in Fig.\,\ref{fig2}d and \ref{fig2}e.
 Our conjecture that the superdomains are formed by the fine stripes of the primary tetragonal ferroelectric domains is illustrated in Fig.\,\ref{fig3}c.
%

%
Let us stress that we have observed that this fine stripe pattern within superdomains  is gradually fading out.
In other words, the blades appear more and more homogeneous. 
Simultaneously, the contrast between adjacent blades observed with the fixed orientation of crossed polarizers is enhanced (see Fig.\,\ref{fig2}c).
In the end, only one system of wide homogeneous parallel stripes persists over the whole sample, as it is schematized in Fig.\,\ref{fig2}f.
These resulting wide stripes obviously correspond to the primary tetragonal domains separated charge domain walls, similar to those shown in Fig.\,\ref{fig1}a.

\section{Theory}

\subsection{Standard WLR theory of the phase front}


Before addressing the properties of superdomain walls and composed precipitates, let us summarize the already known predictions of the
martensitic WLR theory\cite{wechsler1953theory}.
According to our knowledge, the martensitic WLR theory has been first applied to perovskite ferroelectrics in order to explain the formation of the phase front between the paraelectric cubic and ferroelectric tetragonal phase in the seminar work of Ref.\,\onlinecite{didomenico1967paraelectric}. %
Almost in parallel, it has been applied to high-resisitivity BaTiO$_3$ in Ref.\,\onlinecite{Parker_1969}.
Later the theory has been further developed and independently tested or formulated by many others \cite{fesenko1985regularities,fesenko1989domain,dec1993paraelastic,jin2003adaptive,roytburd1993elastic,tagantsev2010domains}.

According to the martensitic theory, the ferroelectric phase transition  proceeds preferentially by propagation of a specific coherent planar phase front, also termed as habit plane, connecting the paraelectric part of the crystal with a suitably twinned ferroelectric part of the crystal. 
In order to ensure the mechanical compatibility between ferroelectric and paraelectric phases, a martensitic twin is formed from two ferroelectric domain states is a suitable volume ratio and acceptable phase fronts are restricted to specific planes with a precisely defined crystallographic orientation. 

Let us consider the usual pseudocubic cartesian axes of the parent perovskite structure,
$m\bar3m > 4mm$ ferroelectric transition and a ferroelectric twin composed of the prevailing tetragonal domain state with polarization along $+z$ and a minority tetragonal domain state with polarization along $+x$.
It is convenient to denote such a twin as [$Z,x$].
According to the theory, the optimal relative volume of the minority state $+x$ in the [$Z,x$] WLR twin, mechanically compatible with the cubic phase, is given by
\begin{equation}
    \xi = (a_{0} -a )/ (c -a ) ~, 
\end{equation}
 where  $ a_{0}$, $a$ and $c$ are lattice parameters of the cubic and tetragonal phase at the point of the phase transition, respectively.
The average spontaneous strain of the representative [$Z,x$] WLR twin, defined with respect to the cubic lattice parameter $a_{0}$, can be expressed as
\begin{equation}
  \epsilon=  \left(
    \begin{array}{ccc}
              0  & 0  & 0 \\
              0  & \varepsilon_{\perp}  & 0 \\
              0  & 0  & \varepsilon_{\parallel} - \xi \Delta ~, 
    \end{array} 
    \right)
    \label{eqn_3x3}
\end{equation}
where $ \varepsilon_{\perp} = (a-a_{0} )/ a_{0} $, $ \varepsilon_{\parallel} = (c-a_{0} )/ a_{0} $,  are spontaneous strain parameters of the primary tetragonal domain state and $\Delta =\varepsilon_{\parallel} - \varepsilon_{\perp}$.
The average polarization of the  [$Z,x$] twin is then
\begin{equation}
    {\bf P} = (\xi, 0, 1-\xi)P_{\rm 0}~,
    \label{eqnP}
\end{equation}
where $P_{0} $ is the equilibrium magnitude of the primary single domain polarization at the phase transition.
In this case, the phase-front unit normal vector, defined as pointing towards the side of the cubic phase, can have four possible orientations
\begin{equation}
{\bf n}= (0, \pm \alpha, \pm \sqrt{1-\alpha^2}) ~,
\label{eqnn}
\end{equation}
where 
\begin{equation}
    \alpha =\sqrt{\frac{- \varepsilon_{\perp}  }{  \Delta (1-\xi)} } ~.
\end{equation}

Interestingly, none of the vectors ${\bf n}$ in eq.\,(\ref{eqnn}) is perpendicular to the average polarization of the WLR twin, listed in in eq.\,(\ref{eqnP}). 
Therefore, the habit plane bears a net bound charge density.
In particular, the head-to-none phase front of [$Z,x$] twin with ${\bf n}= (0, \pm \alpha,  \sqrt{1-\alpha^2})$ posses a positive bound surface charge and the tail-to-none phase front ${\bf n}= (0, \pm \alpha,  -\sqrt{1-\alpha^2})$ surface bears a negative bound surface charge. 
In all cases, the   absolute value of the bound surface charge density on these phase fronts is $\sigma = (1-\xi) \sqrt{1-\alpha^2} P_{0}$.
In fact, it is comparable to the bound charge at the charged ferroelastic domain walls. 

In case of BaTiO$_3$, the volume ratio of primary domain states in the WLR twin is close to 2:1 and the phase fronts are approximately $\{056\}$ crystallographic planes. 
These predictions have been verified by direct observation of such macroscopic phase fronts\cite{fesenko1985regularities,fesenko1989domain,dec1993paraelastic}.

\subsection{Concept of superdomain boundaries}

The observations reported in Figs.\,\ref{fig2} and \ref{fig3} suggest that in the present experiments, conducted under the [$111$]-directed poling field, there are many differently oriented phase fronts and several distinct WLR twin states. 
Therefore, it is reasonable to expect that several such secondary domain states, for brevity denoted as superdomain states, might coexist in the same crystal. 
Distinct superdomains can meet at superdomain boundaries.

%
%
From the point of view of the average strain and polarization, the parent cubic point group symmetry allows 24 equivalent orientational variants of the above described WLR twin states. 
The exact macroscopic crystallographic symmetry\cite{hlinka2016symmetry} implies that these 24 WLR twin states can be considered as 24 domain states of the fully ferroelectric monoclinic species $m\bar3m > m_{+}$ (see species 207 in Ref.\,\onlinecite{hlinka2016symmetry} ). 
Nevertheless, the average strain given by the eq.\,(\ref{eqn_3x3}) distinguishes only 6 distinct average spontaneous superdomain strain tensors.
Therefore, for the considerations of the mechanical compatibility of superdomain walls, it is sufficient to consider the effective ferroelastic  species $m\bar3m > mm_{+}$. 
In other words, for this purpose
it is possible to disregard the sign combinations of the polarization components and consider only the six  ferroelastic superdomain states denoted in the following as Xy, Xz, Yx, Yz, Zx, Zy.

\begin{figure}
    \centering
    \includegraphics[width=0.90\columnwidth]{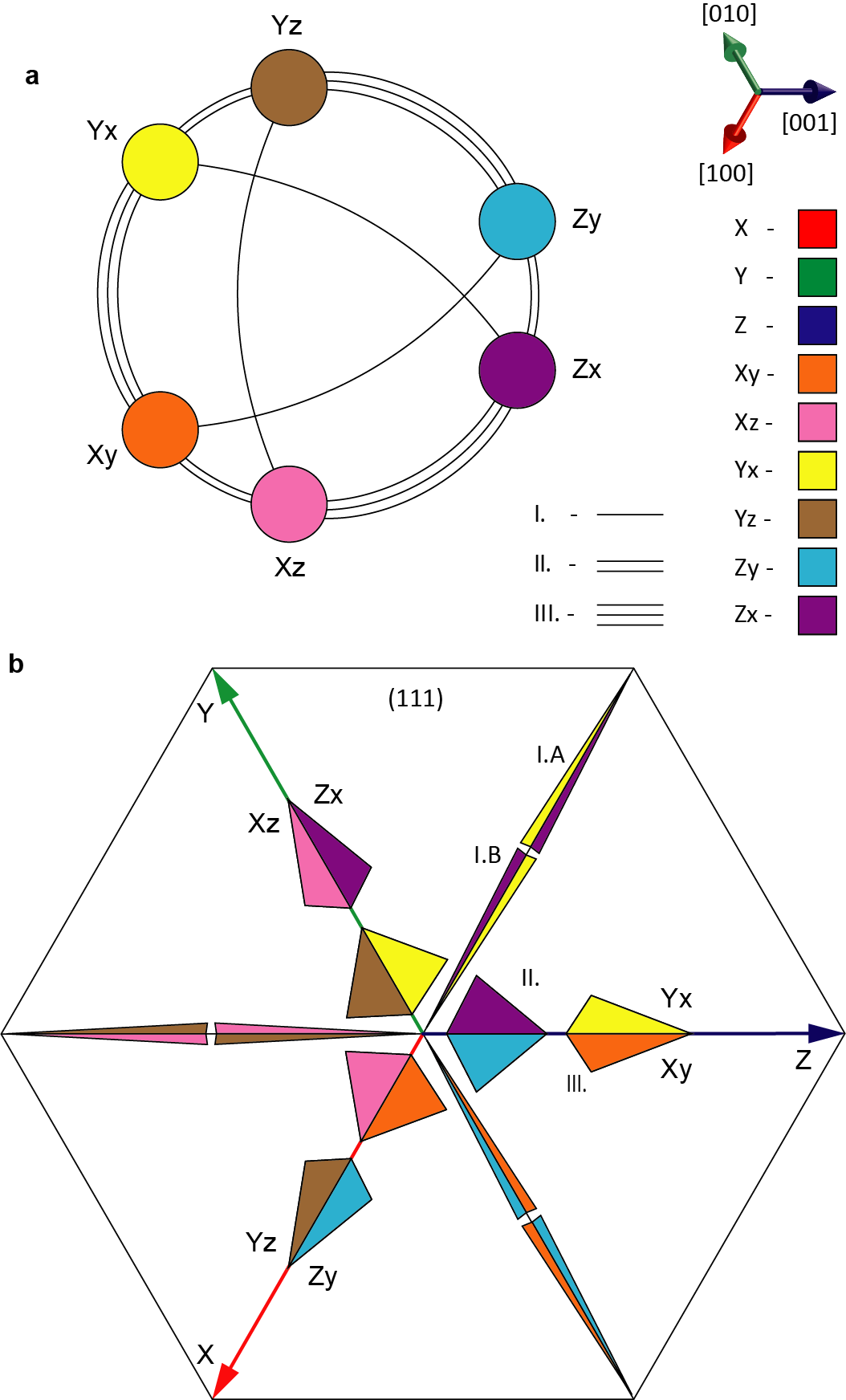}
    \caption{
    %
    %
    Summary of theoretical predictions.
    The vertices of the graph in panel (a) stand for the 6  types of ferroelastic superdomains compatible with the paralectric cubic structure.
    Type of the connecting line (single, double, tripple) indicate the type of the mechanically permissible superdomain boundary of type I., II. and III, respectively. 
    Panel (b) displays the (111) surface sections of the hypothetical polyhedral ferroelectric precipitates delimited from the cubic phase by  mechanically compatible planar phase fronts.
    Same view direction as in the Fig.\,\ref{fig2} is assumed.
    %
    %
    %
    Precipitates are labeled I., II. and III. according to the type of the embedded central superdomain boundary.
    Type I.A and I.B precipitates are composed of two infinite triangular prisms (see Section \ref{2-s}). 
    Type II. and III. precipitates have pentahedral pyramid shape that can terminate by an apex in the volume of the sample. 
    The outer hexagon is merely a guide for eye.
    %
    }
    \label{fig4}
\end{figure}

The orientation of the mutually mechanically compatible superdomain walls among these 6 WLR ferroelastic states was determined from the 
average  spontaneous strain tensors similarly as in  Ref.\,\onlinecite{neuber2018architecture}.
There are three distinct types of mechanically compatible ferroelastic superdomain walls,
labeled as type I., II. and III. (see the graph of Fig.\,\ref{fig4}a). 
Type I. superdomain wall joins two superdomains with a common minoritary primary domain state (for example, superdomain states Xy and Zy). 
Type II. superdomain wall connects two superdomains with a common majority primary state (for example, Xy and Xz) and type III. superdomain wall joins two superdomains having their primary domain population interchanged (for example, Xy and Yx pair). 
Each compatible pair allows two orientations of the boundary, listed in Table I.).
The other pairs of the superdomain states are not mechanically compatible, meaning that no mechanically permissible superdomain walls can be constructed.

\begin{table}[]
    \centering
    \begin{tabular}{c|c|c|c|c|c|c}
        & Xy & Xz   & Yx & Yz & Zx & Zy \\
        \hline
      Xy &
       & (011) & (110)& - & - & (101) \\
       &
       &  (01$\bar{1}$)$^*$ &  (1$\bar{1}$0)$^*$& - & - &  (10$\bar{1}$)$^*$ \\
      \hline
      Xz &
     (011) &  & -& (110)& (101) &-  \\
        &
    (01$\bar{1}$)$^*$ &  & -& (1$\bar{1}$0)$^*$& (10$\bar{1}$)$^*$ &-  \\
     \hline
       Yx &
    (110) & - &  & (101)  & (011) & -  \\
        &
    (1$\bar{1}$0)$^*$& - && (10$\bar{1}$)$^*$ & (01$\bar{1}$)$^*$ & -  \\
    \hline
      Yz &
     - & (110) & (101) & & -& (011) \\
        &
     - & (1$\bar{1}$0)$^*$ & (10$\bar{1}$)$^*$ & & -& (01$\bar{1}$)$^*$ \\
    \hline
      Zx &
    -& (101) &  (011) & - &   & (110) \\
           &
    -& (10$\bar{1}$)$^*$ &  (01$\bar{1}$)$^*$ & - &  & (1$\bar{1}$0)$^*$ \\
    \hline
       Zy &
    (101) & -& -&  (011)& (110) &   \\
           &
    (10$\bar{1}$)$^*$ & -& -&  (01$\bar{1}$)$^*$ & (1$\bar{1}$0)$^*$ &   
    \end{tabular}
    \caption{Orientations of the mechanically compatible superdomain boundaries among the WLR superdomains. For each compatible domain pair, there are two permissible domain wall orientations. One of them, marked by asterisk, is perpendicular to the (111)  surface of the sample.}
    \label{tab:my_label}
\end{table}

\subsection{Termination of superdomain walls at (111) surface}

It turns out that in case of the investigated cubic to tetragonal ferroelectric phase transition, not only the primary ferroelastic walls, but also the secondary ones are $\{011\}$ type planes.
Consequently, these planes are are either perpendicular to the (111) crystal surface, intersecting it  along [$\bar{2}11$], [$1\bar{2}1$], or [$11\bar{2}$] directions,
or they are oblique to the (111) surface, intersecting it along along [$01\bar{1}$], [$1\bar{1}0$], or [$\bar{1}01$] directions.  
The superdomain walls of the former family  are marked by asterisk in Table\,I.
It can be expected, that those with
the orientation perpendicular to the sample are more favorable for reduction of the strain energy in the sample.

\subsection{Termination of habit planes at (111) surface}

By their incidence with respect to the (111) sample surface, the habit planes of the phase fronts can be also divided into two families. 
Members of the first family show the incidence almost perpendicular to the surface, while the others have rather an almost parallel position.
Intersection of the phase fronts  with (111) sample surface obviously falls along the cross product of the phase front and surface normals.
This yields a directional vector with components
($\alpha- \sqrt{1-\alpha^2},\sqrt{1-\alpha^2},-\alpha  $) or another vector symmetrically equivalent to it. 
In this set of vectors, there are 6  directions that are close to [$\bar{2}$11], [1$\bar{2}$1], or [11$\bar{2}$] axes, and these correspond to the phase fronts close to perpendicular to the (111) surface. 
There are also 6 other orientations that are close to [$01\bar{1}$], [$1\bar{1}0$], or [$\bar{1}01$] directions, and these ones correspond to the phase fronts almost parallel to the (111) surface.  
All possible cases are also depicted in Fig.\,\ref{fig4}b.

\subsection{Single-superdomain precipitates}

The canonical WLR scenario of the temperature driven phase transition assumes a single phase front traversing the crystal and leaving a single WLR twin behind. 
In contrast, the  [$111$]-directed poling field used here is expected to favor formation of twinned precipitates of different types.
Moreover, the sides of the sample are typically subject of a lower field, so that the transformation starts in the central part of the sample.
Let us first briefly consider what could be convenient shape of one isolated precipitate formed by a single WLR twin.


%
%
%
%
%

We recall that there are just 2 compatible phase front planes for a fixed
twin, related to the 4 phase front normals listed in eq.\,(\ref{eqnn}).
%
%
It can be anticipated that the mechanically most convenient shape of such a precipitate is a thin platelet, possibly of a lentil-like shape, delimited by only two slightly bend phase fronts, being close to one of the corresponding fully compatible planar orientations. 
It can be expected, that the phase fronts almost perpendicular to the platelet sample surface  will be energetically more convenient than the oblique ones. 
Since the phase fronts on the opposite side of such lentils are necessarily of opposite bound charge, the shape might be systematically somewhat asymmetric.

\subsection{Two-superdomain precipitates}
\label{2-s}

It seems that the above described single superdomain precipitates of the same type would be mutually attracted and merge together,
while the  parallel precipitates delimited by phase fronts in a reversed order of charge densities would be rather electrostatically repulsed from each other, at least before the bound charge is fully screened by mobile electronic and ionic carriers.
In either case, the existence of needle-shape leaf-like motifs documented experimentally in Fig.\,\ref{fig2} and Fig.\,\ref{fig3} clearly suggests that it might be more convenient to form precipitates composed of two different WLR superdomains. 
Guided by the experimental observations of Fig.\,\ref{fig2}, our geometry considerations were further limited to such composed precipitates that are symmetric with respect to the central  ferroelastic superdomain boundary, knowing also from the observations that this boundary is  a $\{110\}$ oriented plane, and that it is perpendicular to the (111) surface. 
In other words, we assumed that the central superdomain boundary orientation is one of those marked by asterisks in Table\,1. 

What could be a convenient shape and crystallographic orientation for such a two-superdomain precipitate?
The simplest 3D objects delimited by planar walls are tetrahedra. 
Therefore, we have been considering precipitates composed by a pair of tetrahedra, each delimited by two phase fronts, a common superdomain wall and the top surface. 
 Let ${\bf n}_1, {\bf n}_2, {\bf w}$ and ${\bf s}$ are  the out-pointing unit vectors perpendicular to the two phase fronts, to the superdomain wall, and to the $(111)$ surface normal, respectively. 
 It can be shown that for a given combination of such normals, the tetrahedron can be formed if
\begin{equation}
  \left[  ({\bf n}_1 \times    {\bf n}_2 ) \cdot  {\bf w}  \right] 
  \left[({\bf n}_1 \times  {\bf n}_2 ) \cdot  {\bf s}  \right] < 0 ~.
\end{equation}
Clearly, if the surface normal is fixed and this tetrahedron formation condition is satisfied for a given combination of ${\bf n}_1, {\bf n}_2 $ and $ {\bf w}$, it cannot be simultaneously satisfied with same ${\bf n}_1, {\bf n}_2 $ but with the opposite $ {\bf w}$.
This implies that a given combination of ${\bf n}_1$ and $ {\bf n}_2$ vectors allows the tetrahedra being formed at only one side of the superdomain boundary. 

It might happen that the vectors fulfill the limiting relation $(  ({\bf n}_1 \times  {\bf n}_2 ) \cdot  {\bf w}  ) =0$. 
Then, the normals ${\bf n}_1, {\bf n}_2 $ and $ {\bf w}$ form an infinite triangular prism instead, which could be limited by the top and bottom (111) surface. 
In this case, another such prism can be constructed with the same ${\bf n}_1$ and $ {\bf n}_2$ pair on the other side of the superdomain pair.

The inspection of available combinations of the phase front normal vectors shows, that if the central superdomain boundary is of type I., then two kinds prismatic precipitates can form by a junction of two distinct triangular prisms. 
The prismatic precipitate of the first kind has its sharp angle phase front wedge pointing along the [$\bar{2}11$], [1$\bar{2}1$]  or [$11\bar{2}$]  direction.  
We denote this case as I.A. 
The second type of the prismatic precipitate has its sharp angle phase front wedge pointing in one of the opposite directions.
It is denoted as the I.B case.
If the central superdomain boundaries are of type II. or III., it is possible to construct a pentahedral precipitate composed of two real tetrahedra.
For a given superdomain pair, the particular crystallograpic orientation of the individual facets of such precipitate can be derived using the Table\,1.
For realistic values of lattice strain, corresponding to the $\{056\}$ type habit planes, the shapes and orientations of these precipitates have been determined. 
The apparent shape of these precipitates given by their termination on the top surface (view from outside along the field direction) is given in Fig.\,\ref{fig4}b.

Nevertheless, it should be stressed that although the precipitates shown in the Fig.\,\ref{fig4}b are formed by the mechanically compatible planar interfaces only, it can be expected that there would be residual strain incompatibility at all their edges.
In reality, only the edges at the relatively sharp wedge-like junctions are likely to be well accommodated by the elastic deformation, while the junctions at obtuse angles are likely to be energetically quite  inconvenient.
Therefore, when possible, the precipitates are likely to join into more complex structures, such as the one shown in Fig.\,\ref{fig2}e, where the junctions with obtuse angles are avoided only the sharp wedges are preserved. 
Note that this peculiar arrangement is formed by an array of alternating head-to-head and tail-to-tail superdomain walls, terminated by two inequivalent zig-zag shaped phase fronts, one of the head-to-none type and one of the tail-to-none type.

\subsection{ Domain states promoted by the electric field}

The electric field applied along the [111] crystallographic direction is favoring the three primary domains states with polarization along +x, +y and +z directions over the three remaining domain states.
Similarly, the ${\bf P.E }$ coupling of the electric field to the polarization should favor superdomain states composed of the +x, +y and +z primary domain states only.
In other words, one expect to the six superdomain states [$X,y$], [$X,z$], [$Y,x$], [$Y,z$], [$Z,x$] and [$Z,y$] to be most likely ones.
Another 6 superdomain states [$X,-y$], [$X,-z$], [$Y,-x$], [$Y,-z$], [$Z,-x$] and [$Z,-y$] could be considered as moderately favored. 
The 6 superdomain states of [$-X,y$] type are unlikely to occur and those of [$-X,-y$] form are probably not formed under the [111] bias field at all. 
Examples of superdomains with different degree of the alignment along the [111]-directed electric field, in particular (a) the strongly favored superdomains [$Z,x$] and [$Y,x$], (b) moderately favored superdomains [$Z,-x$] and [$Y,-x$], (c) moderately disfavored superdomains [$-Z,x$] and [$-Y,x$], and (d) strongly disfavored  superdomains [$-Z,-x$] and [$-Y,-x$].

\begin{figure}[h]
\centering
\includegraphics[width=0.95\columnwidth]{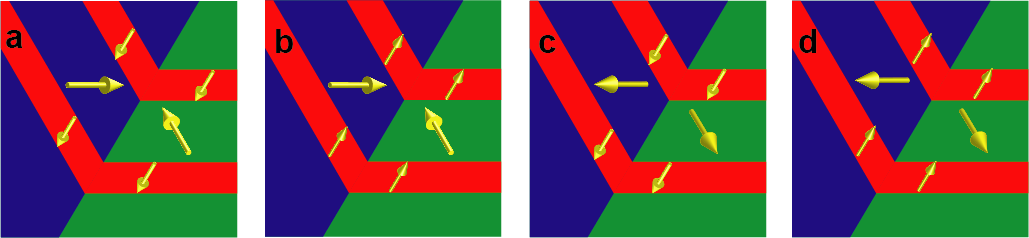}
\caption{
Possible primary domain polarization configurations within a pair of adjacent  martensitic twins (Zx and Yx), sharing a common charged superdomain boundary of type I.B.
 With respect to the field applied along the [111] direction,  panels correspond to a strongly favored (a), moderately favored (b), moderately disfavored (c) and strongly disfavored configurations (d), respectively.
 The arguments in section IV indicate that the transient structure of Fig.\,\ref{fig2}b and Fig.\,\ref{fig3}f is that of the moderately favored case.
%
%
}
\label{fig3bis}
\end{figure}

\subsection{Bound charge at superdomain boundaries}

In general, the ferroelastic superdomain walls may be both charged and uncharged ones. 
Considering that (i) the observed superdomain walls  are perpendicular to the (111) surface, and  that (ii) only the 6 strongly favored superdomains are present in the sample, we are  left with the charged superdomain walls only. 
Let us stress that the same actually holds also if only the weakly favored domain states are considered. 
The restriction of the superdomain type allows to deduce also the charge densities on the
phase front facets of our composed precipitates.
The numerical estimates  of the  bound charge densities at the permissible superdomain walls and at the principal phase front facets is summarized in the Table\,II. 

It is found that the prismatic precipitate of type I.A should contain a charged central tail-to-tail superdomain wall surrounded by two head-to-none phase fronts. 
Conversely, the prismatic precipitate of type I.B should contain a charged central superdomain wall of head-to-head type, surrounded by two tail-to-none phase fronts. 
These conclusions are equally valid for precipitates formed from the weakly favored superdomain states, as well as for precipitates formed from the strongly favored superdomain states.
Otherwise, in case of the strongly field-favored superdomains, one can state that the permissible pentahedral precipitates of type II. have their central superdomain wall of head-to-head type and the phase fronts forming at the (111) surfaces the smaller angles are head-to-none type, while the other pair of phase front facets are of tail-to-none type.
In contrast, the permissible pentahedral precipitates of type III. have their central superdomain walls of tail-to-tail type and the phase fronts are all of head-to-none type. 
 In the case of precipitates formed from the weakly favored superdomain states, the superdomain walls and phase front facets of the pentahedral precipitates of type II. and III. would have the opposite charge densities in comparison to their strongly favored counterparts.

\begin{table}[]
    \centering
    \begin{tabular}{c|c|c|c|c}
    & \multicolumn{2}{|c|}{strogly favored superdomains} &\multicolumn{2}{c}{weakly favored superdomains}  \\
         type &  superdomain wall & phase front & superdomain wall & phase front\\
         \hline
         I.A &  -0.66 $P_0 \sqrt{2}$  & +0.74 $P_0$ & 
         -0.66 $P_0 \sqrt{2}$  & +0.74 $P_0$  \\
         I.B &  +0.66 $P_0 \sqrt{2}$   & -0.74 $P_0$  
         &  +0.66 $P_0 \sqrt{2}$   & -0.74 $P_0$ \\
         II. &  +0.25 $P_0 \sqrt{2}$   & + 0.74 $P_0$   
         &  -0.25 $P_0 \sqrt{2}$   & + 0.74 $P_0$ \\
          III. &  -0.41 $P_0 \sqrt{2}$  & + 0.74 $P_0$
          &  +0.41 $P_0 \sqrt{2}$  & + 0.74 $P_0$
    \end{tabular}
    \caption{Bound charge at principal interfaces of the idealized precipitate forms shown in the Fig.\,\ref{fig4}b. Indicated are values for the central superdomain wall and for phase front facets having a longer intersection with the $(111)$ crystal surface.  Type I.A corresponds to the case with the sharp angle wedge vertex pointing to the $[\bar{2}11]$ direction (in Fig.\,\ref{fig4}b, the outer case), I.B stands for the sharp angle pointing to the opposite direction. Results are listed separately for  strongly and weakly field-favored superdomains, respectively.}
    \label{tab:my_label}
\end{table}

\section{Discussion}

{\it Process of the bulk precipitation under (111) poling field.}
 The above outlined theory allows to draw the most likely interpretation of the principal observations shown in the Fig.\,\ref{fig2}. 
 Upon the reducing the temperature below the paraelectric phase stability limit, the bulk free energy of the tetragonal phase become lower than that of the cubic phase. 
 The transition might be then initiated  by the nucleation and growth of single-superdomain lentil-shaped precipitates, which later coalesce into pairs or larger agglomerates.

Inspired by the observations, we deduce that the initial precipitates most likely have already  the two-superdomain precipitate shape with the central wall of type I. (see Fig.\,\ref{fig2}d and Fig.\,\ref{fig4}). 
 Among others, the other plausible types (type II. and III.) of the symmetric two-superdomain precipitates would have much larger angles of the needle-like blade tips than it is observed in Fig.\,\ref{fig2}b and Fig.\,\ref{fig3}a.
 Such individual two-superdomain precipitates are growing and eventually merging into interconnected arrays of alternating wedges of type I.A and I.B. as suggested in Fig.\,\ref{fig2}e.
 Even more likely, the obtuse edges of the original two-superdomain precipitates surrounded directly by the cubic phase  are serving as natural nucleation spots stimulating growth of the next superdomain wall and the neigbouring superdomain blade. 
Such a chain reaction would best explain the experimental fact that these arrays are effectively growing in their lateral direction as well.
Several additional pieces of evidence for the formation of superdomain boundaries are collected in the Appendix/supplementary material (section VII.)

{\it Conversion of superdomain boundaries into the charged domain walls.}
When the whole volume of the sample turns into the tetragonal phase, the primary domain state ratio in the superdomains  is no more controlled by the mechanical compatibility with the cubic phase.
Therefore, the minority primary domains can gradually shrink  and eventually vanish completely.
In the weakly favored superdomains, this process can be promoted by the applied field.
Since this process of vanishing minority domains is rather systematic, we have concluded that the transient superdomains are indeed only weakly favored by the bias field.
In fact, it is well in line with the reasonable expectation that there is not enough mobile charge in the crystal allowing to compensate the primary domain walls in the superdomains if they would be charge ones.
In other words, the weakly favored superdomains with neutral primary domain walls are preferred over the strongly favored superdomains with charged primary domain walls.
As a result, the superdomains with the partially field-favored content are converted by the [111]-directed field to the primary domain states and the adjacent charged superdomain walls are transformed to the primary charged domain walls with the full bound charge.
For example, once the paraelectric phase is absent, the partially field-favored transient superdomain pair of Fig.\,\ref{fig3bis}b is naturally driven by the [111]-directed field to the 
simple primary twin of Fig.\,\ref{fig2}f. 
A the same time,  the superdomain boundaries are converted to  the regular, macroscopic primary charged domain walls.

 {\it Transport of the charge carriers.}
 The results of Fig.\,\ref{fig2} and its above described interpretation also offers a new insight in the mechanism of the separation of the charge carriers, needed for the compensation of the final distribution of bound charges at charged domain walls. 
 In fact, we can see that at the nucleation stage, the phase fronts are originally very near to the superdomain walls of the opposite charge and in fact, they are directly connected.
 Thus, during the nucleation, the conditions for electronic and ionic mobile charge carriers separation along the boundaries or by the bulk diffusion are certainly much more favorable.
 Subsequently, the compensating charge carriers are probably transported  together with the displacement of the  phase fronts themselves.
 In other words, the motion of the charge carriers is synchronized and governed by the same stimuli that causes the phase fronts to move away from their nucleation sites.
The growing of the charged superdomain walls is also interlinked with the motion of charged phase fronts, because they are both connected by the common wedges.
Since all these boundaries are formally bearing sizeable bound charge, we believe that it is likely that they are conductive and the charge transfer between the adjacent superdomain walls of the opposite polarity might be facilitated by a possible electric conduction along the superdomain walls and the phase front zig-zags themselves.
However, the detailed theory of the transport of compensation charge carriers is beyond the scope of this work.


{\it Primary domain wall content of superdomains.}
In principle, all derivations made so far are independent on the type of the primary ferroelastic  domain walls in the superdomains.
Obviously, the primary ferroelastic  domain walls can be either charged or neutral ones.
In general, the energy of the neutral primary domain walls are much lower than that of the charged ones.
Therefore, the neutral primary walls seems to be more likely to be formed.
Interestingly, those that are observed in the Fig.\,\ref{fig2} clearly belong the family of $\{011\}$ planes, perpendicular the (111) direction.
This orientation is certainly favorable for decreasing of the overall  elastic energy of sample with a (111) oriented thin platelet shape.
Nevertheless, this observation is leaving us with two competing options for the interpretation of the type of the primary domain wall type. 
The first possibility is that the superdomains of Fig.\,\ref{fig2}b are the strongly favored ones.
Then, the primary domain walls observed there must be the charged domain walls.
The second possibility is that these observed superdomains are the moderately favored ones.
Then, the involved primary domain walls are the neutral ones. 
Energetically, the superdomain with neutral primary domain walls seems to be more plausible.

\section{Conclusion} 

In summary, by performing in-situ observations of the domain formation during the frustrative poling of BaTiO$_3$ single crystals near the ferroelectric phase transition, we have found multiple evidences for the formation of charged superdomain boundaries. 
These superdomain boundaries are formed already in the early stage of the phase transformation if not simultaneously with the formation of the charged tetragonal-cubic phase fronts themselves.
While the charged phase fronts necessarily perform translation motion trough the crystal in order to suppress the residual cubic phase until the transformation is completed, the superdomain boundaries are relatively immobile and they grow mainly in their length.
Nevertheless, translation motion of phase fronts and extension of superdomains is interlinked as the phase fronts are directly attached to the superdomain boundaries. 

In the case of the volume nucleation scenario, which was the most efficient one in the sense of the formation of multiple charged domain walls, the final charged ferroelastic domain walls were shown to be directly derived from these superdomain boundaries.
The triple junction of a superdomain boundary and two adjacent phase fronts form a narrow wedge, that probably plays an essential role in facilitating the transport of the electronic and ionic charge carriers during the propagation of these charged interfaces throughout the crystal.
The observed self-organized tertiary superdomain structures strongly suggest an efficient chain formation mechanism consisting in nucleation of the new superdomain walls while keeping a sharp angle termination at the backside of the last narrow wedge triple junction in the band.

The ensemble of observations was successfully interpreted in terms of the  WLR theory, extended to describe the multiple superdomain structures and composed precipitates. 
In particular, the theoretical analysis based on the mechanical compatibility of WLR superdomains allowed to understand the specific habitus and the fine structure of the ferroelectric precipitates in the observed transient domain patterns.
Similar theoretical framework might be applied  also to other domain engineered states of polar perovskites, for example in the broad family of relaxor ferroelectrics.
We believe that the insights in the mechanism of the charged domain formation in BaTiO$_3$ materials, which reveal the fundamental role of the superdomain walls, will trigger a new interest in domain engineering perspectives of tailoring the novel nanoscale based properties of ferroic materials.
 
 
 

\section{Experimental Section}

Single crystal samples used in this work and their domain structures have been carefully prepared using the methods described in detail elsewhere\cite{bednyakov2015formation}. The principal experiments of this work presented in Figs.\,\ref{fig1} and \ref{fig2} were made using crystals with a high effective mobile charge density about $10^4$ C/m$^3$.
For complementary observations shown in Fig. S1 we have used samples of originally stoichiometric crystals substantially reduced in oxygen deficient atmosphere in order to achieve an effective charge density of $10^2-10^5$ C/m$^3$. 
Finally,  the restricted charged domain formation reported in Figs. S2, S3a and S3c have been investigated in   
nominally pure stoichiometric dark yellow crystals with effective charge density of $10^2-10^3$ C/m$^3$.
Charge density in all crystals was estimated for the temperature 573K.
All materials were supplied by GB group http://www.crystalsland.com.
For in-situ observations, both main surfaces were polished to a 1 micron quality and covered with 10\,nm thick transparent platinum electrodes that allowed both to engineer and to observe the domain structure by optical means.
 The optical polarizing microscopy was operated using both reflected and transmitted mode and polarized and nonpolarized light.
Observations were performed at room temperature as well as during the cooling from the paraelectric to the tetragonal phase using the heating/cooling stage equipped with electric leads allowing application of the poling electric voltage.
The poling field up to $1.2$kV/mm was applied along the viewing $[111]$ direction using the high voltage DC power supply SRS PS325.

\section{Acknowledgements}

We acknowledge the support from the Czech Grant Agency (GACR project No.20-05167Y). Both authors are appreciating stimulating discussions about the subject with Prof. A. Tagantsev from EPFL Laussane.
Authors also appreciate the support of Prof. N. Setter from EPFL Laussane who initiated the investigation in this field under the EU 7th Framework Programme (FP7/2007–2013)/ERC grant agreement n [268058].

\section{Conflict of Interest}

The authors declare no conflict of interest.

\section{Data Availability Statement}

The  data  that  support  the  findings  of  this  study  are  available  from  the  corresponding author upon reasonable request.

\bibliographystyle{naturemag}
\bibliography{ReferencesLess}

\end{document}


\title{Supplementary information to the manuscript "Charged domain walls in BaTiO$_3$ crystals emerging from superdomain boundaries"}



\author{Petr S. Bednyakov}
\author{Ji\v{r}\'{\i} Hlinka}
\affiliation{FZU-Institute of Physics$,$ The Czech Academy of Sciences$,$ Na Slovance 2$,$ 18221 Praha 8$,$ Czech Republic}
\email{bednyakov@fzu.cz, hlinka@fzu.cz}



\date{\today}

\maketitle

This supplement contains description of additional observations supporting the presence of charged superdomain walls in (111)-poled BaTiO$_3$ crystal plates.

\begin{figure}[h!]
\centering
\includegraphics[width=0.85\columnwidth]
{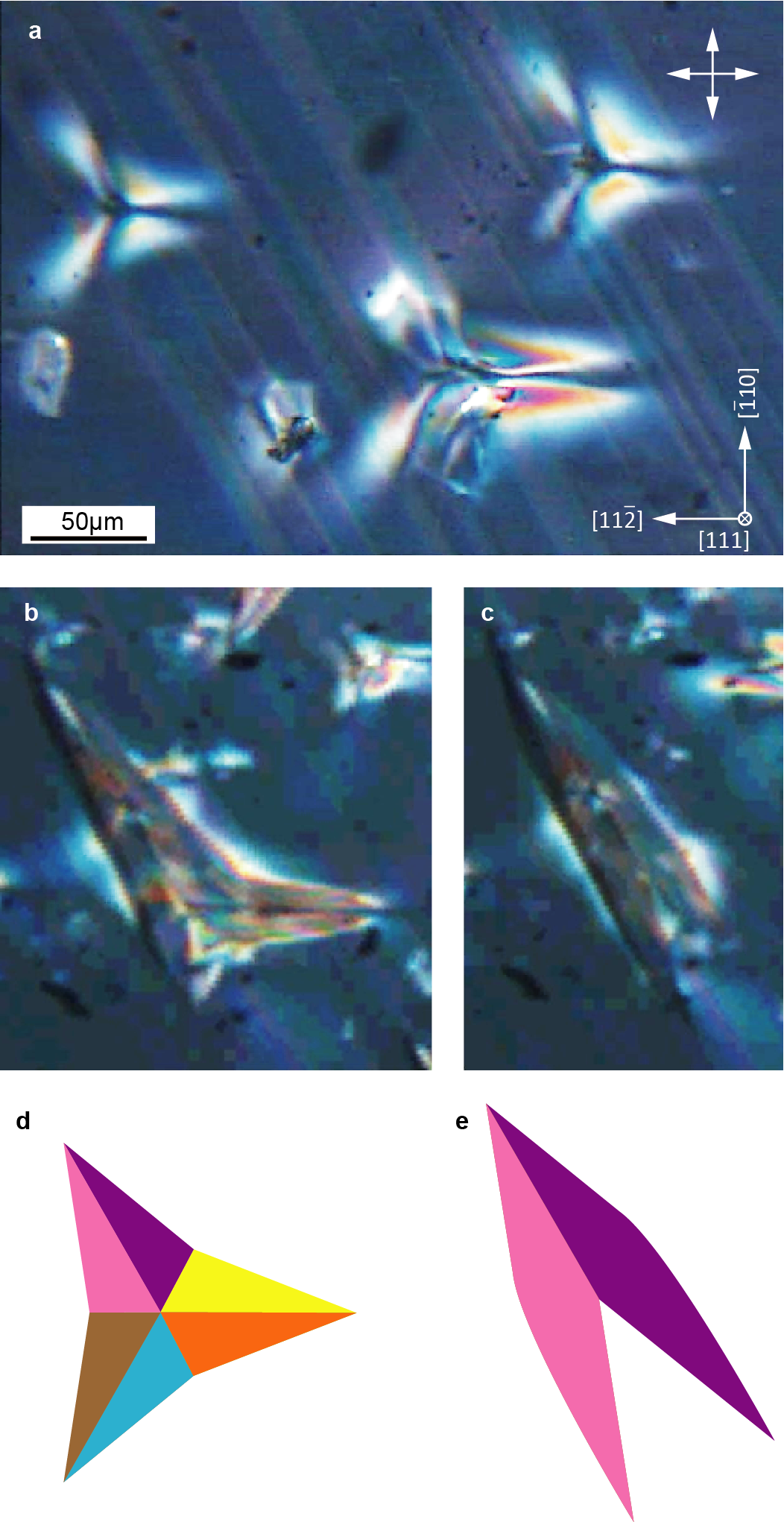}
\caption{Formation of nuclei at the surface of the crystal.
Surface nuclei are typically formed at local defects located at the surface of the samples with a sufficiently conductive surface layer.
Positions of surface nuclei show a marked butterfly birefringence pattern,
showing strong local strain gradients present even before the borders of the nuclei are apparent (a). 
In this case the nuclei often expands arms in more than a single direction
(b).
In all cases each arm appears to be composed by two blades, suggesting the important role charged superdomain walls. 
Panels (d) and (e) suggest a plausible idealized structure of the surface nuclei considering the crystallographic orientation of type III. pentagonal nuclei of Fig.4. 
}
\label{figS1}
\end{figure}

\section{ Surface nucleation}
 Obviously, the details of the nucleation mechanism may vary depending on various specific conditions under which the phase transition is induced.
 For example, some of our originally less conductive samples were treated in the oxygen reducing atmosphere in order to increase the conductivity of the sample.
 Since the treatment is inducing more defects in the near surface layer of the sample, it can be expected that surface nucleation of the ferroelectric phase is enhanced in such crystals. 
 Moreover, structural surface defects and the resulting surface nuclei in such sample also substantial local strain gradients, as evidenced by the characteristic birefringence butterfly patterns seen in the optical image of one such samples (Fig.\,\ref{figS1}a). 
 By comparison of the apparent angles of the surface image of the already grown nuclei with that of the theoretical predictions summarized graphically in Fig.4, we concluded that the observed nuclei involve type III. central superdomain boundary.
 In some cases, several nuclei are connected together (see Fig.\,\ref{figS1}b), again most likely in order to avoid the mechanically incompatible obtuse wedges.
 For example, the nuclei of Fig.\,\ref{figS1}b reminds a fragment of hypothetical star-like pattern  composed of all six superdomain types (see Fig.\,\ref{figS1}d). 
 In other cases, the nuclei was showing an arrow type geometry, possibly similar to the geometry sketched in Fig.\,\ref{figS1}e. 
 However, there was no tendency towards the concerted organization of the nuclei into a more ordered domain structure type like in the case of bulk nucleation described above. 
 In fact, it turned out that this particular sample and the related poling process scenario was not very efficient in creation of multiple stable charged domain walls.
 Instead, it either yielded the so-called zigzag domain walls that we shall briefly address later on, or the resulting charged domain walls were so unstable that they fully decomposed after the removal of the poling electric field.
 
\section{Edge nucleation}
\begin{figure}[h]
\centering
\includegraphics[width=0.95\columnwidth]
{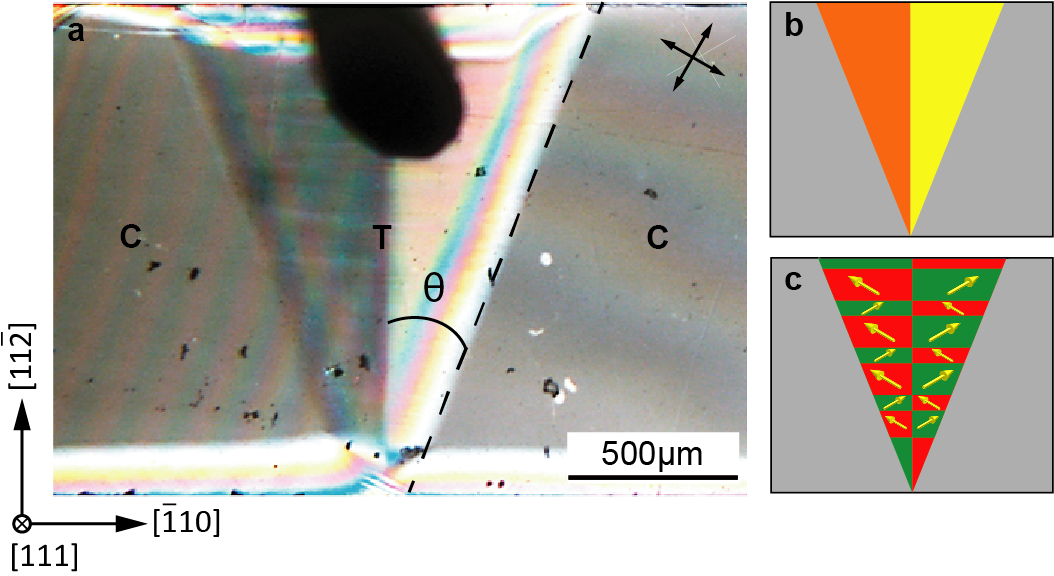}
\caption{
Charged superdomain wall formed at the edge crystal.
(a) Optical micrograph showing the transient wedge domain, separated in two parts by a clear boundary, identified a superdomain wall as schematically shown in the panel (b).
The angle between the superdomain wall and phase front boundary ${\theta\approx21^\circ}$ corresponds best to that of the type III. wedge geometry of Fig.4. 
Dark spot on the top of the picture is a contact used for the electric field application,                            
the top and the bottom part of the image corresponds to the opposite edges of the sample.
Labels "T" and "C" stands for the tetragonal and cubic phases, resp.,
crossed directors shows the orientation of the polarizer and analyzer.
The primary domain structure pattern  (c) at the sample surface can be deduced from the direction of the observed stripes, mechanical compatibility and the assumption that the majority primary domains are oriented favorably to the applied field.
}
\label{figS2}
\end{figure}

 Another interesting phase transformation scenario was observed when the tetragonal phase nucleated at the edge of the crystal.
 Fig.\,\ref{figS2} shows  such transient wedge domain imaged at about 1 K below $T_{\rm start}$. 
 This sample was a (111)-oriented, 100 micron thick plate made of a crystal with a rather low conductivity of the order of 0.01-0.1 S/m.
 The nucleation site was most likely directly related to the local electric contact. 
 Again, by comparison with the theoretical shapes of two-superdomain precipitate images on the (111) sample surface, we can assign it  as the wedge domain with the central superdomain of type III. 
 In this case, the fine primary domains could be seen to have the orientation indicated schematically in Fig.\,\ref{figS2}c.
They are not very clear on the micrograph because these walls are oblique to the main surface. 
 Assuming again that the primary domain walls are electrically neutral ones, we can conclude that here the involved superdomains formed  are the strongly field-favored ones.
 Obviously, the resolution of the optical image was not sufficient to determine the microscopic structure of the superdomain wall. 
 %
 %
 %
Still, we can conclude that the overall bound charge at this superdomain wall should be negative (in other words, it has a tail-to-tail orientation of the  average polarization).
At the same time, the phase fronts are of head-to-none type, implying that the negative mobile charge carriers are needed to screen these moving phase fronts. 
 In principle, this poling strategy can be suitable when formation of just a single charged domain wall would be desired.

\begin{figure}[h]
\centering
\includegraphics[width=0.9\columnwidth]
{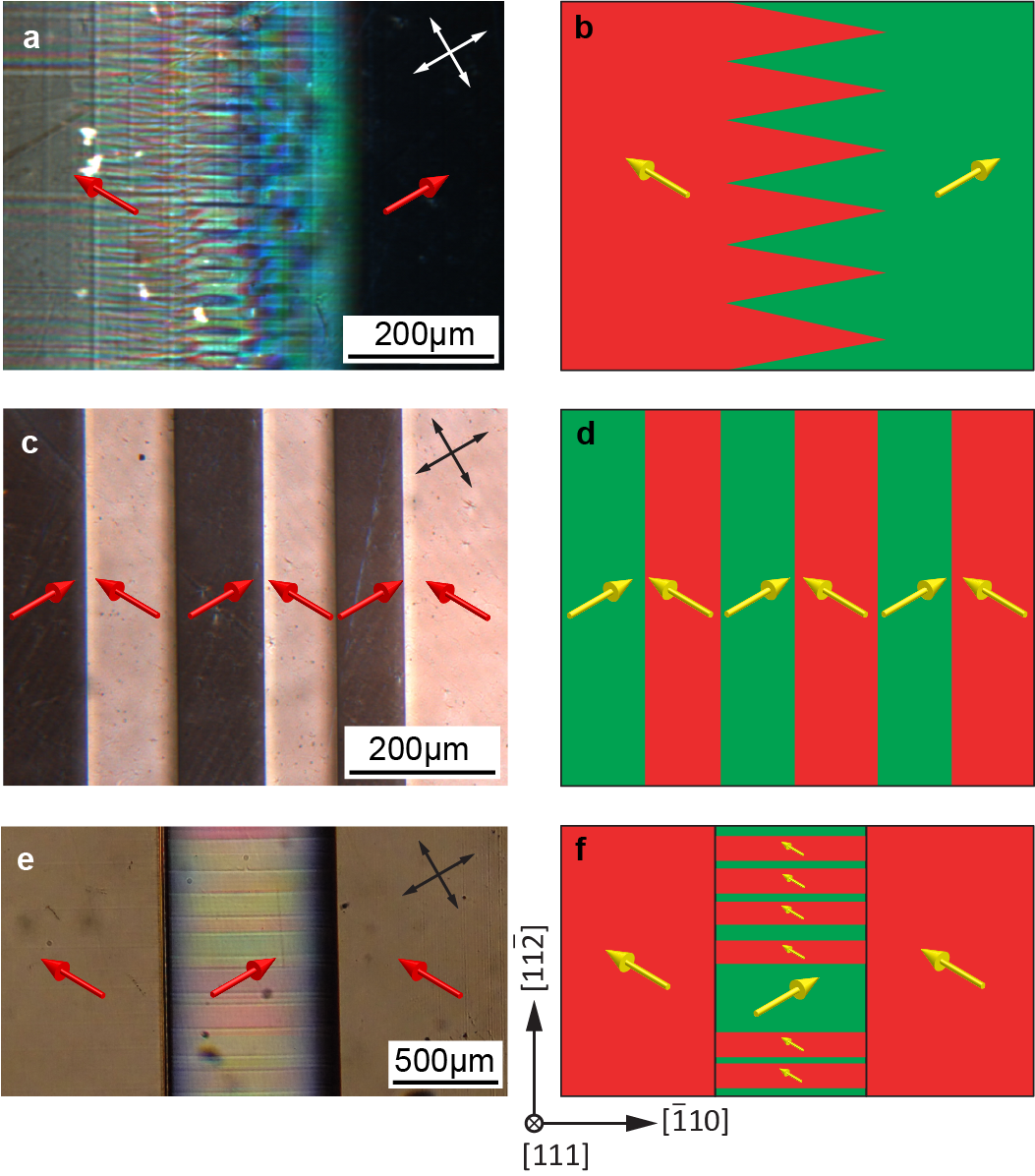}
\caption{
The three basic examples of the final charged domain walls of the same macroscopic crystallographic orientation as observed in various $(111)$-oriented BaTiO$_3$ crystal plates. 
(a) Tail-to-tail zigzag charged domain wall composed of nearly neutral domain walls and its schematic drawing (b); an array of ideal planar charged domain walls (c) and their schematic drawing (d); (e) charged superdomain walls separating a superdomain region from the two primary domains and its schematic drawing (f).
Crossed black/white arrows in right-upper corner of panels (a,c,e) correspond to the orientation of polarizers.
%
%
}
\label{figS3}
\end{figure}

\section{ Zig-zag domain walls}
%
It has been repeatedly observed that frustrative poling of low conductivity BaTiO$_3$ samples results in formation of the so called zigzag charged domain walls, in particular in case of tail-to-tail charged walls that require charge compensation by the less available positive mobile carriers. 
On a macroscopic scale, such a zig-zag domain wall (example shown in Fig.\,\ref{figS3}a) can be rationalized as a charged domain wall, differing from the normal one by its considerable thickness.
However, on the microscopic scale, it is clearly formed by a zigzag array of the almost neutral primary domain walls that are just slightly tilted from their ideal neutral wall orientation.
Due to this small tilt, such primary walls carry a small bound charge, which, however, in projection to the macroscale domain wall orientation gives the usual bound charge density value. 
In the present case, the nominal  orientation of the neutral primary walls involved in the zigzag are already inclined with respect to the sample surface.
Therefore, the  zigzag image is partly distorted.
Nevertheless, by counting the number of the zigzag turns per given length it can be estimated that the angle between adjacent zigzag lamellae is less than 10  degrees.  
This corresponds to about at least 10 times smaller bound charge density per surface area of the primary microscopic wall, or, equivalently, it means that the surface of the zigzag microscopic wall is at least 10 larger than that of the corresponding macroscopic wall (or of its ideal planar counterpart, such as the one in Fig.\,\ref{figS3}c,d). 
Interestingly, the macroscopic width of the zigzag wall (the length of its tooth) is of the order of the typical final distance between charged domain walls themselves or even larger (see Fig.\,\ref{figS3}c).
In this sense, the zig-zag itself can be considered as a sort of secondary domain, although here the volume ratio of the two primary domain states is probably 1:1 and we could not see any tight relation to the WLR superdomains discussed previously.
We can speculate that the effective width of the zig-zag wall can be related to the volume needed to collect the required amount of charge, while  the  width of the zig-zag tooth could be perhaps related to the characteristic distance from which the compensation charge can be collected by the usual diffusion processes. 
%
 
\section{ Superdomain wall between primary and secondary domains.}
Alternatively to zig-zag walls, in some cases we have observed  secondary domain walls connecting primary domains and the twinned secondary domains, such as in Fig.\,\ref{figS3}e. 
Here the primary fine stripe  domains in the twinned central area are most likely the neutral ones, in other words, those that
are inclined with respect to the viewing direction. 
The sharp borders between the dark and light area in the photograph Fig.\,\ref{figS3}e thus correspond to the charged macroscopic tail-to-tail and head-to-head charged domain walls, respectively. 
Because of the twinning in the central area, the overall bound charge density is reduced in comparison with the primary charged walls  of Fig.\,\ref{figS3}c.
Schematic interpretation of this structure is proposed in Fig.\,\ref{figS3}f. 
Note that all large-area planar interfaces in this domain structure are mechanically compatible ones.
The experiment does not reveal the microstructure the superdomain wall, simplest assumption is that it formed by an aligned system of wedge domains
ended on a common plane. 
In either case, on the macroscopic scale, the superdomain wall of Fig.\,\ref{figS3}f requires smaller overall amount of screening charge then the primary domain of Fig.\,\ref{figS3}d.

In principle, we may speculate that the secondary (twinned) domain in the center of Fig.\,\ref{figS3}e,f is probably residuum of a WLR superdomain state,
 fully favored by the bias electric field, which was surrounded by two superdomains only partially favored by the bias electric field. 
 As we have argued  already, it can be expected that in the course of the domain structure coarsening, the bias field will promote the transformation of the  partially favored superdomains into the primary domains.
 On the other hand, there is no such obvious driving force present in case of  the fully favored  superdomains, and therefore the process of elimination of the fine stripe primary domains is expected to be much slower, hindered or completely frozen.

 
 

\bibliographystyle{naturemag}
\bibliography{ReferencesLess}